\documentstyle[preprint,aps]{revtex}
\begin{document}
\preprint{OITS-540}
\draft
\title{Proton Life-Time Problem In Finite Grand Unified Theories\footnote{Work
supported in part by the Department of Energy Grant No. DE-FG06-85ER40224} }
\author{N.G. Deshpande, Xiao-Gang He and E. Keith}
\address{Institute of Theoretical Science\\
University of Oregon\\
Eugene, OR 97403-5203}
\date{April, 1994}
\maketitle
\begin{abstract}
We study proton decay in finite supersymmetric SU(5) grand unified theories. We
find
that the dimension-five operators due to color triplet higgsino
induce too rapid a proton decay. This behaviour can be traced to the large
Yukawa
couplings to the first generation that are necessary for finiteness.
\end{abstract}
\newpage
Proton decays are predicted in many grand unified theories (GUTs)\cite{gut}.
Experimentally
no proton decays have been observed\cite{bound}. The stringent experimental
bounds on proton
decays can provide interesting constraints on GUTs\cite{pd,ano,hisano}. It has
been shown that in the
minimal supersymmetric (SUSY) SU(5) model, a large region in parameter space
can be ruled out from the consideration of proton decays\cite{ano,hisano}. In
this paper we study
proton decays in a class of finite SUSY GUTs, namely the finite SUSY SU(5)
models.
We show that models considered so far are ruled out by experimental
bounds on the proton life-time.

There have been many studies of finite
GUTs\cite{finite,raby,span,discrete,list}. This is a class of interesting
GUTs. It supports strongly the hope that the ultimate theory does not need
infinite renormalization.  In order to have a finite theory to all orders, the
$\beta$ functions for the gauge coupling and Yukawa couplings have to be zero
to all orders.
The requirement that the $\beta$ function of the gauge coupling be zero greatly
restricts the allowed matter representations in a
theory once the gauge group is given. The $\beta$ function of the Yukawa
couplings being zero can put additional constraints on the theory.
A list of possible finite theories are given in Ref.\cite{list}.  A
particularly interesting class of theories are the ones based on the
$SU(5)$ gauge group with supersymmetry. There are several solutions satisfying
the requirement that the $\beta$ function of the gauge coupling be zero.
However,
the requirement that the $\beta$ function for the Yukawa couplings be zero
further restricts the allowed solutions.
If one requires that $SU(5)$ is broken by the Higgs mechanism to $SU(3)_C\times
SU(2)_L\times U(1)_Y$ with three generations of matter fields,  only one
solution is allowed with
$5$, $\bar 5$, $10$, $\bar {10}$ and  $24$ chiral multiplets with
multiplicities (4,7,3,0,1)\cite{list}. This model contains one 24 ($\Sigma$) of
Higgs for the SU(5) breaking, $4(5 +\bar 5)$ $(H_\alpha\;,\; \bar H_{\alpha})$
of Higgs some of which will be used for electroweak breaking and the remaining
$3(\bar 5 +10)$ are identified with
the three generation matter fields. With this content, the most general
superpotential that may be written, consistent with renormalizibility, $SU(5)$
invariance and R-parity conservation is of the form
\begin{eqnarray}
W = q Tr\Sigma^3 + M Tr\Sigma^2 + \lambda_{\alpha\beta} \bar H_\alpha \Sigma
H_{\beta} + m_{\alpha\beta}\bar H_\alpha H_\beta + {1\over 2}g_{ij\alpha}
10_i 10_j H_\alpha + \bar g_{ij\alpha} 10_i \bar 5_j \bar H_\alpha\;,
\end{eqnarray}
The indices $\alpha,\; \beta$, and $i,\;j$ run from 1 to 4 and 1 to 3,
respectively.

The requirement that the $\beta$ functions for the Yukawa couplings are zero at
the one-loop level implies,
\begin{eqnarray}
\Sigma: &{189\over 5}q^2 = 10g^2 -
\lambda_{\alpha\beta}\lambda^{\alpha\beta}\;,\nonumber\\
\bar H_\alpha: & \bar g_{ij\alpha} \bar g^{ij\beta} = {6\over 5}
(g^2\delta^\beta_\alpha - \lambda_{\alpha\gamma}\lambda^{\beta\gamma})\;,
\nonumber\\
\bar 5_i: & \bar g_{ki\alpha}\bar g^{kj\alpha} = {6\over 5} g^2\delta^j_i\;,\\
H_\alpha: & g_{ij\alpha} g^{ij\beta} = {8\over 5}(g^2\delta^\beta_\alpha -
\lambda_{\gamma\alpha}\lambda^{\gamma\beta})\;,\nonumber\\
10_i: & 2 g_{ik\alpha} g^{jk\alpha} + 3 g_{ik\alpha} g^{jk\alpha} = {36\over 5}
g^2\delta^j_i\;.\nonumber
\end{eqnarray}

This set of equations constrains the allowed values for the Yukawa couplings.
However it is not restrictive enough so that all fermion masses and
Kobayashi-Maskawa (KM) mixing angles can be predicted. The requirement of
all-loop finiteness may
further reduce the parameters. In Ref.\cite{discrete}
imposing an additional $Z_7\times Z_3$ symmetry, a unique solution to eq.(2) is
found
\begin{eqnarray}
g^2_{111} = g^2_{222} = g^2_{333} = {8\over 5}g^2\;,\nonumber\\
\bar g^2_{111} = \bar g^2_{222} = \bar g^2_{333} = {6\over 5}g^2\;,\nonumber\\
\lambda_{44}=g^2\;,\;\;\;\;\; q^2 = {5\over 21}g^2\;.
\end{eqnarray}
All other tri-linear couplings are zero. This is a very interesting theory
because it
is an all-loop finite theory\cite{discrete}.
All the Yukawa couplings are expressed in terms of gauge coupling $g$. This
allows
one to further predict some fermion masses.

In the above model only $H_4 (\bar H_4)$ can develop vacuum expectation values
in order that the doublet-triplet mass splitting is possible for the doublets
which
break $SU(3)_C\times SU(2)_L\times U(1)_Y$ to $U(1)_{em}$. Since the quadratic
coupling $m_{\alpha\beta}$ is diagonal due to the $Z_7\times Z_3$ discrete
symmetry
and $H_4(\bar H_4)$ do not couple to quarks and leptons, the fermions are all
massless. This
problem can be solved by softly breaking the $Z_7\times Z_3$ discrete symmetry
with off-diagonal $m_{\alpha\beta}$ entries. In this way one can find solutions
such that each Higgs doublet can develop a vacuum expectation value and at the
same time it is still possible to maintain the doublet-triplet mass splitting.
All fermions can now have masses\cite{span}. Below the unification scale the
model is effectively the same as the minimal SUSY standard model. Adding soft
breaking terms, one can get supersymmetry breaking. Since the theory is
spontaneously broken, the finiteness conditions do not restrict its
renormalization properties. Carrying out the
renormalization group analysis, the top quark mass has an upper-bound of
190 GeV. If the Higgs doublets which develop VEV only couple to the third
generation, the top quark is determined to be between $175$ to $190$
GeV\cite{discrete}.

There are, however, several problems with this model. Because the Yukawa
couplings are
diagonal, all KM angles are zero. This is  inconsistent with experiments. This
problem can be solved by abandoning the
diagonal solution to eq.(2). It is possible to find a solution of eq.(2)
such that
KM matrix can be reproduced. A possible solution is
\begin{eqnarray}
\bar g_{ij\alpha} = \sqrt{6\over5}g(\delta_{i,1}\delta_{\alpha,1}
+\delta_{i,2}\delta_{\alpha,2}+\delta_{i,3}\delta_{\alpha,3})V_{ij}
\end{eqnarray}
with all other couplings the same as in eq.(3). Here $V_{ij}$ is the KM matrix.
This model
has the same predictions for the quark masses. It does not satisfy the
discrete $Z_7\times Z_3$ symmetry.

There is another problem related to fermion masses in this model,
as has been noted in Ref.\cite{raby}. Because there are only $5$ and $\bar 5$
Higgs representations to generate masses for quarks and charged leptons, this
model also predicts
the wrong mass relations for the first two generations:
$m_e=m_d$, $m_\mu = m_s$ at the GUT scale. This is a common problem for
SU(5) models with only 5 and $\bar 5$ Higgs representations to generate fermion
masses. If higher dimension operators
are somehow allowed, this problem can be solved. For example,
adding a $(10\times \bar 5)
(\Sigma \bar H_\alpha)$ term can correct the wrong mass relations. However,
this solution
is not consistent with the finiteness conditions. This, however, is not the
major
problem. In the following we will show that even if we relax the conditions to
allow the above additions
to the theory, the model has another problem. It predicts too rapid a proton
decay.

There are several mechanisms by which proton decays
may be induced in SUSY SU(5) theories. The exchanges of heavy gauge bosons is
one. In the finite theory discussed here the contributions from heavy gauge
bosons are the same as in the minimal SUSY SU(5). The proton decays due to this
mechanism have been extensively studied\cite{pd,hisano,fpd}, and  can easily
satisfy the experimental lower bounds\cite{hisano,fpd}. In the minimal SUSY
SU(5) model, exchange
of scalar color triplets will also generate dimension-six operators which can
mediate proton decays. There, however, due to small Yukawa couplings, the decay
rates
due to this mechanism is much smaller than the contribution from the heavy
gauge bosons. In the finite SUSY SU(5), the
Yukawa couplings are of the same order of magnitude as the gauge coupling. The
scalar color triplets induced proton decays are comparable with the heavy gauge
boson contributions, and can easily satisfy the
experimental bound because the scalar color triptlet masses are also of the
same
order of magnitude as the gauge bosons and could even be somewhat heavier.  The
most significant contributions to the proton decays come from the
dimension-five operator induced by exchanging color triplet higgsinos $H_C$ and
$\bar H_C$ of
$H_\alpha$ and $\bar H_\alpha$\cite{pd,ano,hisano}. In the minimal SUSY SU(5)
model, this mechanism is the dominant one and considerably restricts
the allowed region in parameter space of the model\cite{hisano}. In the
finite SU(5) model, experimental bounds on proton decays all but make these
models unacceptable.

The diagrams for the dimension-five induced four-fermion operator responsible
for proton decays are shown in Fig. 1.  There are other similar contributions
that arise by replacing the chargino $\tilde w$ by a gluino, or a zino. It has
been argued that the dominant ones are from chargino
exchange\cite{pd,ano,hisano}, and we shall only need to consider chargino
constributions. The four-fermion baryon number violating effective Lagrangian
at 1 GeV can be written down explicitly as\cite{hisano}
\begin{eqnarray}
L &=& {\alpha_2\over 2\pi M_{H_C^\alpha}} g_{ii\alpha} \bar g_{kk\alpha}
V_{jk}^*A_SA_L\nonumber\\
&&\times [ (u_id'_i)(d'_j\nu_k) ( f(u_j,e_k) + f(u_i, d'_i))+
(d'_iu_i)(u_je_k)(f(u_i,d_i)+f(d_j', \nu_k))\\
&&+ (d'_i\nu_k)(d'_iu_j)(f(u_i, e_k)+f(u_i, d'_j))+(u_id'_j)(u_ie_k) (f(d'_i,
u_j)+f(d'_i,\nu_k))]\nonumber\;,
\end{eqnarray}
where $d'_i = V_{il}d_l$; $f(a,b)= m_{\tilde w}[m^2_{\tilde a} \ln (m^2_{\tilde
a}/m^2_{\tilde w})/(m^2_{\tilde a}-m^2_{\tilde w}) - (m_{\tilde a} \rightarrow
m_{\tilde b})]/( m^2_{\tilde a}-m^2_{\tilde b})$ is from the loop integral, and
$m_{\tilde a, \tilde b}$ are the s-fermion masses,
$A_S \approx 0.59$, $A_L \approx 0.22$\cite{hisano} are the QCD correction
factors for the
running from $M_{GUT}$ to SUSY breaking scale and from SUSY breaking scale to 1
GeV, respectively, and the Yukawa couplings are evaluated at 1 GeV.

Because all $g_{ii\alpha}$ and $\bar g_{jj\alpha}$ are equal in the model we
are considering, the dominant contributions to the proton decays
will be the ones involving only particles in the first generation. The dominant
baryon number violating decay modes are:
$p\rightarrow \pi^+ \bar \nu_e$, $p\rightarrow \pi^0 (\eta) e^+$,
$n\rightarrow \pi^0 (\eta) \bar \nu_e$, $p\rightarrow \pi^- e^+$.

Finally to
obtain the life times of the proton and neutron, we employ the chiral
Lagrangain
approach to parametrize the hadronic matrix elements\cite{chiral}.
We have
\begin{eqnarray}
\Gamma&(&p\rightarrow \pi^+ \bar \nu_e) = 2\Gamma(n\rightarrow \pi^0 \bar
\nu_e)
= \beta^2 {m_N\over 32\pi f^2_{\pi}}|C(duu\nu_e)(1+D+F)|^2\nonumber\;,\\
\Gamma&(&n\rightarrow \eta \bar \nu_e)
=\beta^2 {(m_N^2-m_\eta^2)^2\over 64\pi f^2_{\pi}m_N^3}3|C(duu\nu_e)(1-{1\over
3}(D-3F))|^2\;,\\
\Gamma&(&n\rightarrow \pi^- e^+) = 2\Gamma(p\rightarrow \pi^0 e^+)
= \beta^2 {m_N\over 32\pi f^2_{\pi}}|C(duue)(1+D+F)|^2\nonumber\;,\\
\Gamma&(&p\rightarrow \eta e^+)
=\beta^2 {(m_N^2-m_\eta^2)^2\over 64\pi f^2_{\pi}m_N^3}3|C(duue)(1-{1\over
3}(D-3F))|^2\;,\nonumber
\end{eqnarray}
where $D = 0.81$ and $F=0.44$, which arise from the strong interacting
baryon-meson
chiral Lagrangian, $f_\pi = 132$ MeV is the pion decay constant, and $m_N$ and
$m_\eta$ are the neucleon and $\eta$ meson masses, respectively. The parameter
$\beta$ is estimated to be in the range $0.03$ GeV$^3$ to $0.0056$
GeV$^3$\cite{beta}. The parameters
$C(duu\nu)$ and $C(duue)$ are the coefficients of the operators
$(du)(u\nu)$ and $(du)(ue)$ which can be read off from eq.(5). We have
\begin{eqnarray}
C(duu\nu_e) &=& {4\alpha^2_{em}\over \sin^4\theta_W}{\bar m_b \bar m_t\over
m_W^2 \sin 2\beta_H}{A_SA_L\over
M_{H_C^1}}V_{ud}^2V^*_{ud} (f(u,e)+f(u,d))\;,\nonumber\\
C(duue) &=&{4\alpha^2_{em}\over \sin^4\theta_W}{\bar m_b \bar m_t\over m_W^2
\sin 2\beta_H}{A_SA_L\over
M_{H_C^1}}V_{ud}V^*_{ud} (f(u,e)+f(u,d))\;.
\end{eqnarray}
In the above we have used $g_{111}\bar g_{111} = g_2^2\bar m_b\bar
m_t/m_W^2\sin2\beta_H$ as a good approximation. Here the quark masses are at 1
GeV. $\tan \beta_H$ is the ratio of the vacuum expectation value of $H_1$ to
that of $\bar H_1$. It is predicted to be about 50. The top quark mass at 1 GeV
$\bar m_t$ is about 470 GeV\cite{discrete}. Using these values,
we obtain the partial life-times for some of the baryon number violating decays
as
\begin{eqnarray}
\tau&(&p\rightarrow \pi^0 e^+)\approx \tau(n\rightarrow \pi^0\bar \nu_e)
\approx 6 \times 10^{17}\times P \;\;years\;,\nonumber\\
\tau&(p&\rightarrow \pi^+\bar \nu_e)\approx \tau(n\rightarrow \pi^- e^+)
\approx 3 \times 10^{17}\times P\;\; years\;,\\
\tau&(&p\rightarrow \eta e^+)\approx \tau(n\rightarrow \eta \bar \nu_e)
\approx 2 \times 10^{18}\times P\;\; years\;,\nonumber
\end{eqnarray}
where
\begin{eqnarray}
P = \left ({0.003\;GeV^3\over \beta}\right )^2\left (
{M_{H_C}\over 10^{17}\; GeV}{TeV^{-1}\over f(u,d)+f(u,e)}\right )^2\;.
\end{eqnarray}

The value $\beta = 0.003 GeV^3$ is at the lower end of the estimations. The
color triplet higgsino mass can not be too much larger than $10^{17}$ GeV. Even
if we allow it to be the same order as the Planck mass, these partial
life-times are in contradiction with experiments if the factor
$I=TeV^{-1}/(f(u,d)+
f(u,e))$ is of order one. In the model discussed above, there are no other
possible sources to cancel the above contributions. The only possible way out
is to have a very small $I$. If all s-fermion and
chargino masses are of order TeV, the factor $I$ has to be of O(1). If chargino
is much heavier than s-fermions, $f \approx (\ln (m_{\tilde w}^2
/m_{\tilde f}^2))/m_{\tilde w}$. In order to satisfy the experimental bounds on
the
partial life-times, the mass of the chargino has to be in the
$10^8\; TeV$ region for $m_{H_C} = 10^{17}$ GeV. If the s-fermion masses are
much larger than
the chargino mass, $f \approx m_{\tilde w}/m_{\tilde f}^2$. In this case, the
s-fermion masses have to be larger than $2\times 10^3$ TeV for $m_{\tilde w} >
100$ GeV
and $m_{H_C} = 10^{17}$ GeV.
All these solutions require that SUSY be broken at a scale much much larger
than
a few TeV. However such solutions spoil the nice feature of solving the
hierarchy problem that is the rationale for using SUSY theories in the first
place. This high scale may also cause problem for the correct predictions of
$\sin^2 \theta_W$. From these considerations, the model discussed above is
either ruled out, or quite unattractive needing a large SUSY breaking scale.

In order to solve the proton life-time problem, one needs to find solutions to
the finiteness conditions such that the Yukawa couplings for the first
generation are much smaller than the gauge coupling. This is not ruled out but
it may be difficult to find such  an all-loop finite theory. We expect this
problem to arise in most finite thoeries of grand unification that allow proton
decay.

\newpage

\begin{picture}(100,200)(10,20)

\put(100, 150){\line(1,0){200}}
\put(170, 150){\line(0,-1){80}}
\put(230, 150){\line(0,-1){80}}
\put(100, 70){\line(1,0){200}}
\put( 120, 160){\makebox(0,0){$Q_L$}}
\put( 270, 160){\makebox(0,0){$L_L$}}
\put(120, 60){\makebox(0,0){$Q_L$}}
\put(270, 60){\makebox(0,0){$Q_L$}}
\put(240, 110) {\makebox(0,0){$\tilde H_C$}}
\put(230, 110) {\makebox(0,0){$\times$}}
\put(160,110){\makebox(0,0){$\tilde w$}}
\put(170,110){\makebox(0,0){$\times$}}
\put( 200, 160){\makebox(0,0){$\tilde Q_L$}}
\put(200, 60){\makebox(0,0){$\tilde Q_L$}}
\put(200, 30){\makebox(0,0){(a)}}

\put(100, -20){\line(1,0){200}}
\put(170, -20){\line(0,-1){80}}
\put(230, -20){\line(0,-1){80}}
\put(100, -100){\line(1,0){200}}
\put( 120, -10){\makebox(0,0){$Q_L$}}
\put( 270, -10){\makebox(0,0){$L_L$}}
\put(120, -110){\makebox(0,0){$Q_L$}}
\put(270, -110){\makebox(0,0){$Q_L$}}
\put(240, -60) {\makebox(0,0){$\tilde w$}}
\put(230, -60) {\makebox(0,0){$\times$}}
\put(160,-60){\makebox(0,0){$\tilde H_C$}}
\put(170,-60){\makebox(0,0){$\times$}}
\put( 200, -10){\makebox(0,0){$\tilde L_L$}}
\put(200, -110){\makebox(0,0){$\tilde Q_L$}}
\put(200, -140){\makebox(0,0){(b)}}

\end{picture}

\vspace{6cm}
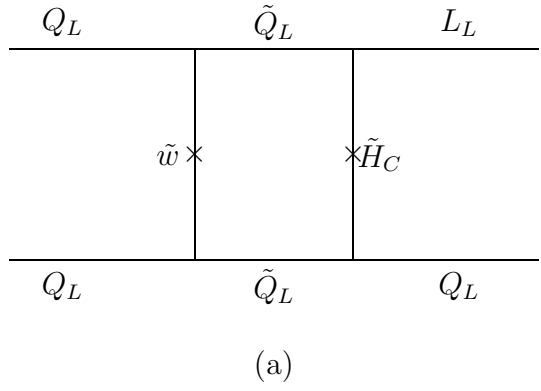
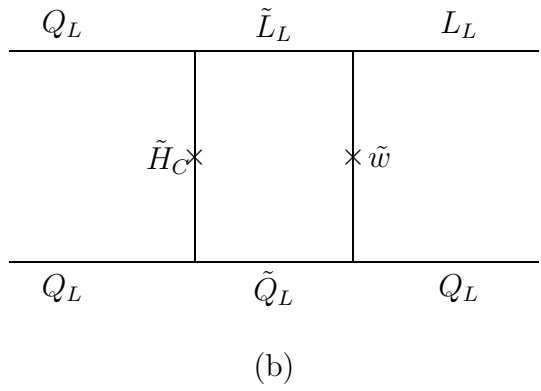
Fig. 1. Dimenison-five proton decay operator due to color
triplet higgsino exchange.

\end{document}